\documentclass{iau} 
\usepackage{graphicx}
\usepackage{xcolor}
\usepackage{url}
\usepackage[colorlinks,urlcolor=teal]{hyperref}
\usepackage[percent]{overpic}

\title[] 
{The potential of Ca~II~K observations \\ for solar activity and variability studies
}

\author[Ilaria Ermolli et al.]   
{Ilaria Ermolli$^1$, Theodosios Chatzistergos$^2$, Natalie A. Krivova$^2$ \and Sami K. Solanki$^{2,3}$}

\affiliation{$^1$INAF -- Osservatorio Astronomico di Roma, 
	Monte Porzio Catone, Italy 
	$^2$Max-Planck-Institut f\"ur Sonnensystemforschung, 
G\"ottingen, Germany  \\[\affilskip]
	$^3$ School of Space Research, Kyung Hee University, 
Republic of Korea}

\pubyear{2018}
\volume{xxx}  
\jname{Title of your IAU Symposium}
\editors{A.C. Editor, B.D. Editor \& C.E. Editor, eds.}
\begin{document}

\maketitle

\begin{abstract}
Several observatories around the globe started regular full-disc imaging of the solar atmosphere in the Ca~II~K line in the early decades of the 20$^{th}$ century. These observations are continued today at a few sites with either old spectroheliographs or modern telescopes equipped with narrow-band filters. The Ca~II~K time series  are unique in representing long-term variations of the Sun's chromospheric magnetic field. However, meaningful results from their analysis require accurate processing of the available data and robust merging of the information stored in different archives. 
This paper provides an  overview of the  historical and modern full-disc Ca~II~K observations, with focus on their quality and the main results obtained from their analysis over the last decade.

\keywords{Sun: activity, plages, Sun: chromosphere, Sun: magnetic fields}
\end{abstract}

\firstsection 

\section{The Ca~II~K line}
The visible spectrum of the Sun, the brightest star in our sky,  contains millions of absorption lines. 
Two of the deepest and broadest such lines are the resonance doublet lines of singly-ionized calcium at 3933 \AA~and 3968 \AA. First observed by Joseph von Fraunhofer in 1814, these lines were designated as K and H, respectively,  in his catalogue of prominent absorption features.  \cite[Linsky \& Avrett (1970)]{Linsky1970}  consider  that ``The significance of these lines lies in the fact that they are the only resonance lines in the visible spectrum from the dominant stage of ionization in the upper photosphere and lower chromosphere of an abundant element. 
Consequently, these lines are very opaque with line centre optical depths of order 10$^7$ in the photosphere and 10$^4$ at the temperature minimum. The entire cores of the lines are thus chromospheric and can serve as convenient probes of the structure and physical properties of the chromosphere.''

\begin{figure}[t]
\begin{center}
\includegraphics[height=4.5cm,trim=2cm 11cm 1.8cm 9cm,clip=true]{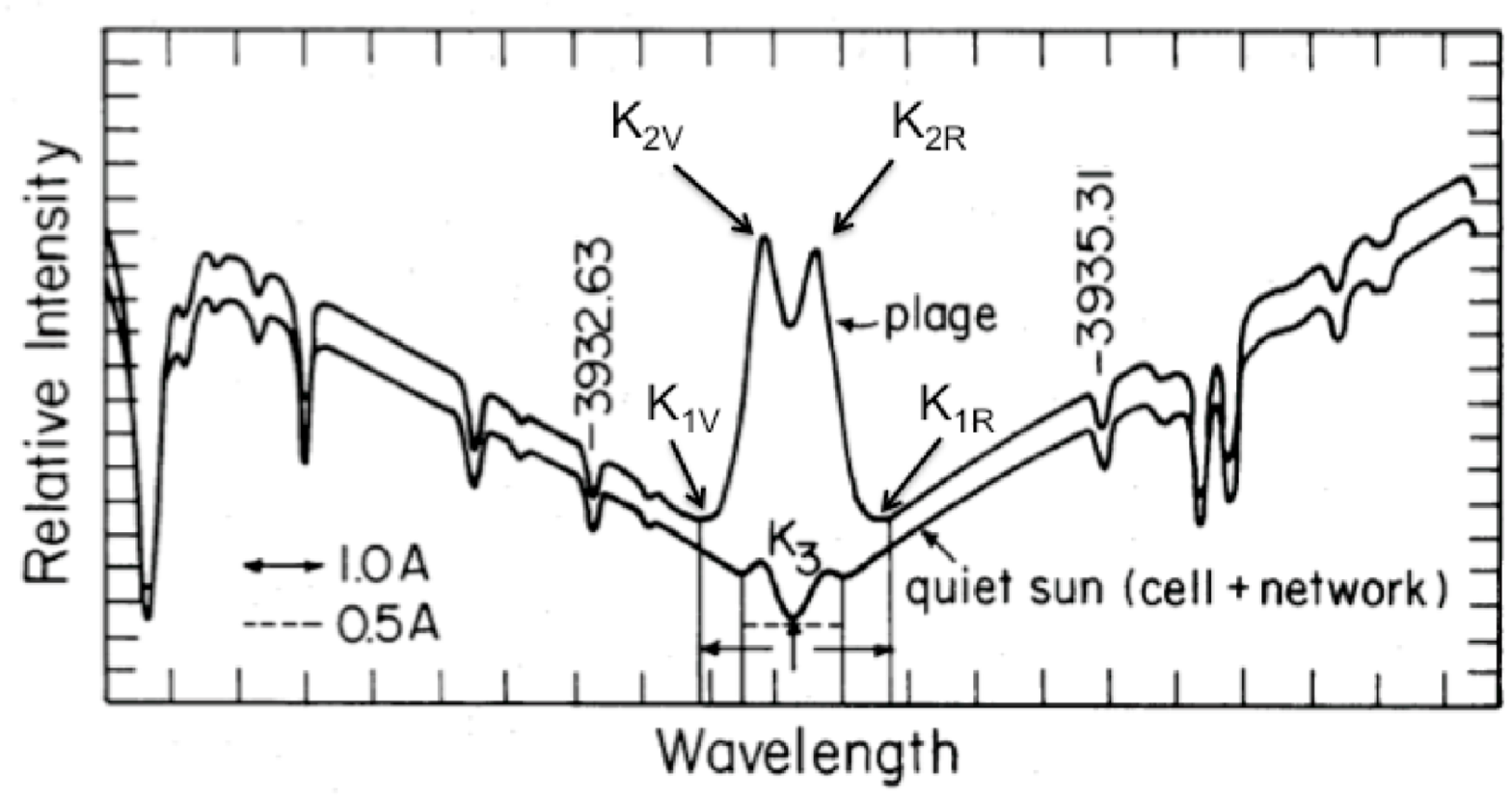} 
		\caption{Ca~II~K line profile for quiet Sun (lower curve) and plage region (upper curve). Adapted from \cite[Skumanich et al. (1984)]{Skumanich1984}.}
		\label{fig2}
	\end{center}
\end{figure}

Figure 1 
shows two spatially-averaged intensity profiles of the Ca~II~K line measured in a quiet Sun and a plage region superimposed on each other. The line profile is characterized by the two peaks K$_{2V}$ and  K$_{2R}$ towards the Violet or Red part of the spectrum relative to the line centre, respectively, the reversal at the line centre K$_3$, and the two secondary minima K$_{1V}$ and K$_{1R}$.  These line features result from sampling of different parts in the inner solar atmosphere. 
From the line wings towards the line centre, the line samples the photosphere with increasing height until the temperature minimum, where the K$_1$ minima are formed. The decreasing temperature with height in the photosphere produces a regular absorption line. Closer to the line core between the K$_1$ minima, the line samples the low chromosphere where the temperature increases with height, the source function essentially follows the Planck function and the line thus gets brighter with increasing height. This is the case nearly up to the atmospheric heights where the K$_2$ peaks are formed, where the source function becomes smaller than the Planck function and absorption dominates again, causing the K$_3$  minimum. These line features appear qualitatively the same for both the quiet Sun and the plage region, but for the latter the K$_2$ peaks and the K$_3$ minimum display a significant increase of the contrast relative to the quiet region. This shows that the chromospheric heating is greatly enhanced in regions where the magnetic field is stronger.


It is worth noting that all individual line features  introduced above occur within  1~\AA~interval.  
\cite[Ermolli et al. 2010]{Ermolli2010} studied the intensity response function for Ca~II~K observations taken with different filters and showed that for data taken with a 1~\AA~bandwidth filter centred at the K$_3$, 58--82\% of the contribution at disc centre is from atmospheric  heights below 500 km. For a 2.5~\AA~wide filter, i.e. similar to that of the present-day 
HINODE SOT/BFI (\cite[Tsuneta et al. 2008]{Tsuneta2008}) and 
broad Ca~II~K Rome-PSPT (\cite[Ermolli et al. 2007]{Ermolli2007}) observations, this contribution is as high as 84--94\%. Besides, for observations with a 1~\AA ~bandwidth filter centred in the red wing of the Ca~II~K line,  more than 99.6\% of the contribution is from atmospheric  heights below 160 km.

\section{Overview of historical and modern observations}

In the early 1890s, Henri Alexandre Deslandres and George Ellery Hale  independently put into practice an instrument  to register full-disc photographic images of the chromosphere at the Ca~II~K and other spectral lines. 

This new instrument, called spectroheliograph, comprises two main parts. The first part is a system consisting of either one or two mirrors directing the Sun's light to a an objective lens that focusses the image of the Sun at the entrance window of the second part of the instrument. This includes a prism or diffraction grating and a monochromator to allow the passage of  a single wavelength through the exit window of the instrument. Both the entrance and exit windows are slit shaped. After the exit window  the photographic material used to record the solar image, most commonly being a photographic plate, or film, and more recently CCD cameras, is placed. A proper displacement of the objective lens allows the full-disc image of the Sun to pass in front of the instrument's entrance slit. With the motion  of the Sun's image at the entrance slit, a proper motion of the photographic material (or of the recording CCD) allows a full-disc image of the Sun to be recorded at the wavelength the instrument is operated. 

Following the first two models at the Paris and Kenwood observatories, several spectroheliographs were built and put in regular operation at different sites, e.g. in 1904 at the Kodaikanal, in 1915 at the Mt Wilson, in 1917 at the Mitaka, in 1926 at the Coimbra, and in 1931 at the Arcetri observatories. The spectroheliographs at the Coimbra and Paris observatories, each equipped with a 45 cm coelostat and 25 cm objective lens, are still operated to this day. In 2008 and 2002, respectively, the photographic plates were replaced with CCD cameras. 

In the 1960s, regular full-disc Ca~II~K observations started  at several sites  also with  telescopes equipped with narrow band  Lyot-type and interference filters, e.g in 1964 at the Rome, 1968 at the Kandilli, and in 1971 at the Big Bear observatories. 

A large fraction of the historical Ca~II~K observations is still stored in its original form, i.e., in  photographic archives, some of which have recently  been digitized, e.g. the Arcetri (1931--1974),  Kodaikanal (1907--2007), Kyoto (1926--1969), Mitaka (1917--1974), and Mt Wilson (1915--1985) ones.

At present, routine  full-disc Ca~II~K  observations are performed with modern instruments at the Baikal, Brussels, Chrotel-Teide, Kanzelh\"ohe,  Mitaka, Pic du Midi, Rome, San Fernando, and Vala\v{s}sk\'{e} Mezi\v{r}\'{i}\v{c}\'{i} Observatories. These series 
started in 2003, 2012, 2009, 2010, 2015, 2007, 1996, 1986, and 2013, respectively. 

Further information on existing Ca~II~K archives can be found in e.g. \cite[Chatzistergos (2017)]{chatzistergos_analysis_2017} and \cite[Chatzistergos et al. (2018a)]{chatzistergos_analysis_2018a}.

\section{Scientific programs}

Since early observations, regular  Ca~II~K data  have been used to monitor the evolution of the most prominent disc features, in particular of plages and prominences.
Besides, they have long served as diagnostics of the solar chromosphere. For example, Ca~II~K data have been used to understand 
the formation of the Ca~II~K line and to produce an atmosphere model of the broad height range  sampled by the line. Indeed,  the temperature structure of  the  semi-empirical atmosphere models by \cite[Vernazza et al. (1981)]{Vernazza1981} and \cite[Fontenla et al. (1993)]{Fontenla1993} were also tested on the  available Ca~II~K observations. 
Over time, analysis of the differences between spatially-averaged profiles and high-resolution observations  of the Ca~II~K  line in different disc regions have also given hints to the strongly  inhomogeneous and dynamic nature  of the chromosphere. 
However, the most prominent aspect of the Ca~II~K line is its ability to provide information about solar magnetism, due to the large increase in the intensity of the  K$_3$ line core when sampling bright magnetic regions (plage). The potential of the Ca~II~K observation to serve as proxy of the magnetic field, which was first noticed by  \cite[Babcock \& Babcock (1955)]{Babcock1955}, has inspired many scientific programs aimed at determining the evolution of the solar magnetic field. To this aim,  the position, dimensions, number, type, and evolution of the different magnetic regions observed on the solar disc were derived from the observations collected at several sites. The measurements of these various parameters were tabulated, graphically represented, and published in volume series by  several observatories. Copies of the photographic observations were also produced and distributed to various research centres. Finally, the connection between the Ca~II~K emission and magnetism has long served to study the magnetic activity of stars other than the Sun. 

The photometric  observations obtained over the last two decades   
have also been used to analyse the radiative properties of the different disc features  in time. For example, \cite[Ermolli et al. (2007)]{Ermolli2007} studied the photometric properties of plages  over the activity cycle, from analysis of Rome-PSPT data obtained from 1998 to 2005, showing that selection effects associated with identification methods can produce significant differences in the results. They also showed that the plage contrast is  a function of the selection method  and the heliocentric angle, but also of the feature size, the activity level, and the content of the analysed images. Besides, \cite[Ermolli et al. (2010)]{Ermolli2010} 
analysed the radiative emission of various types of solar features, such as quiet Sun, enhanced network, plage, and bright plage regions, studying  the dependence of the measured radiative emission on the filter bandpass and comparing results from observations  with those derived from spectral synthesis performed on semi-empirical atmosphere models in the literature. 

Over the last decade, digitization of some prominent photographic archives has initiated extensive exploitation of the Ca~II~K series for studies of the long-term variation of the chromospheric magnetic field and a variety of retrospective analyses. 
To date, historical Ca II data have been mainly used to produce e.g. time series of the disc coverage of plage and network regions, i.e. of the other obvious magnetic features on the solar disc beside sunspots. 
For example,   \cite[Ermolli et al. (2009b)]{Ermolli2009b}, \cite[Foukal et al. (2009)]{Foukal2009}, \cite[Tlatov et al. (2009)]{Tlatov2009}, and \cite[Chatterjee, Banerjee \& Ravindra (2016)]{Chatterjee2016} presented plage area time series derived from analysis of different archives with distinct image processing methods. The annual mean time series presented by these authors show some similarities between independent analyses, but also display significant differences in the overall long-term trends and rank ordering of activity cycle amplitudes. Besides, 
since Ca~II~K observations track the spatial distribution of magnetic flux distribution over the disc, \cite[Harvey (1992)]{Harvey1992}, \cite[Ermolli et al. (2009)]{Ermolli2009}, \cite[Chatterjee, Banerjee, Ravindra (2016)]{Chatterjee2016} produced butterfly diagrams of plage regions identified on the Ca~II~K  observations of the Mt Wilson, Arcetri, and Kodaikanal observatories, respectively. These diagrams complement similar data obtained from sunspot observations and extend butterfly diagrams derived from magnetograms to times when such data were not available. Moreover, to study  the magnetic field at different heliographic latitudes and the toroidal and poloidal components of the solar cycle, \cite[Sheeley, Cooper \& Anderson (2011)]{Sheeley2011} and \cite[Pevtsov et al. (2016)]{Pevtsov2016} produced Carrington maps from Ca~II~K observations of  the Mt Wilson and Kodaikanal archives. 

Finally, an analysis of the supergranulation scale variation based on present-day Rome-PSPT Observations and historical Kodaikanal data was presented by e.g. \cite[Ermolli et al. (2003)]{Ermolli2003} and \cite[Chatterjee et al. (2017)]{Chatterjee2017}, respectively. Network area measurements were presented  by \cite[Foukal \& Milano (2001)]{Foukal2001} and \cite[Priyal et al. (2014)]{Priyal2014}.

Understanding the variation of plage and network properties from analysis of Ca~II~K observations has a broader impact than just describing two of the most prominent solar disc features. Indeed, such measurements  may be useful for constraining models of magnetic  flux transport on the surface of the Sun during multiple sunspot cycles, and for estimating the long-term contribution of facular emission to the total and spectral solar irradiance. Accurate and significant analysis of long-term Ca~II~K series has thus a lot of potential for new knowledge in solar physics.

\section{Comparison among series}

\begin{figure}[t]
	\begin{center}
		\includegraphics[height=8cm,trim=0.5cm 7.3cm 0.5cm 7.3cm,clip=true]{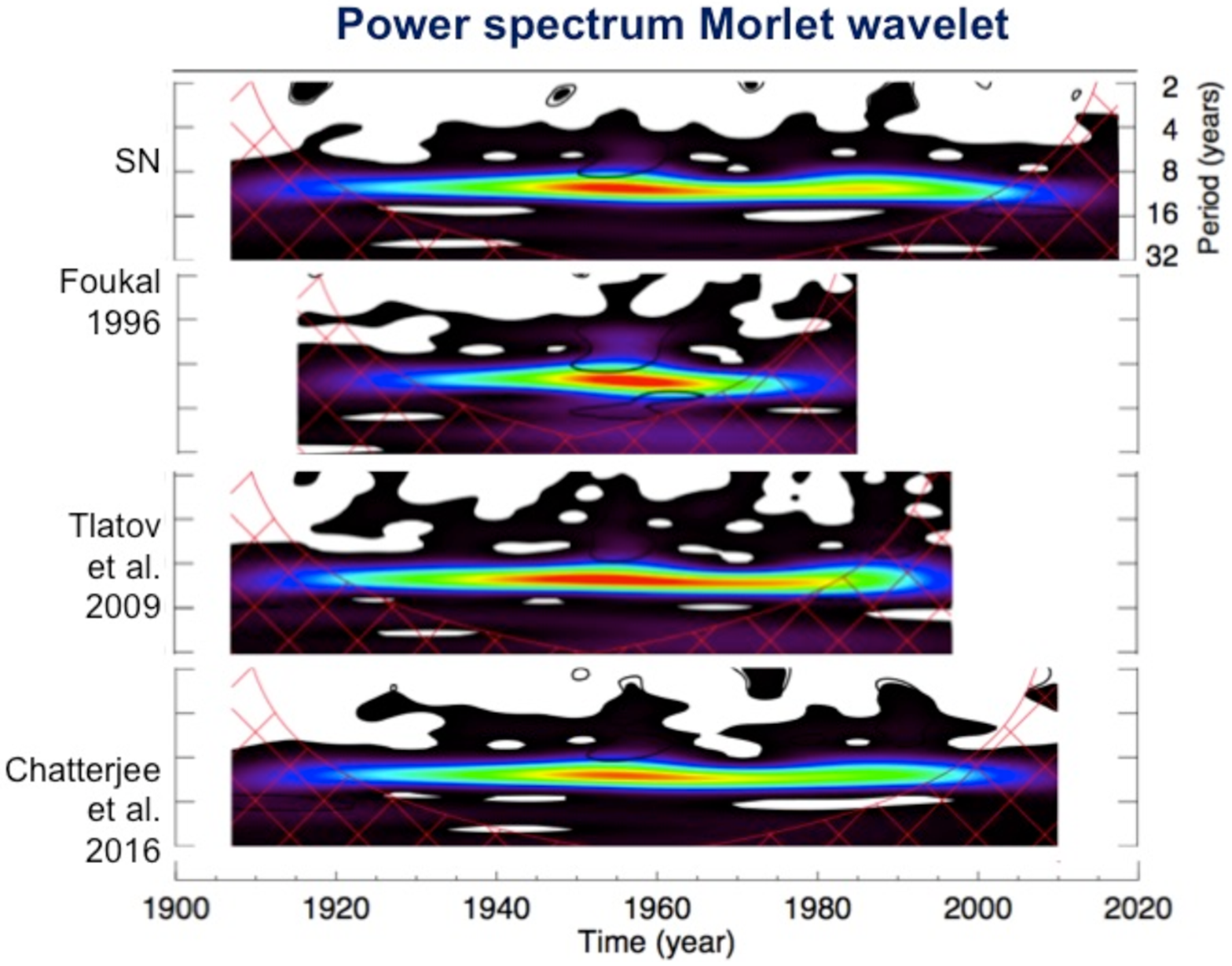} 
		\caption{Example of the power spectrum  of plage area timeseries reported in the literature  using the Morlet wavelet. From top to bottom, power spectrum of the sunspot number series (SN) and of the plage area values by \cite[Foukal (1996)]{Foukal1996}, \cite[Tlatov et al. (2009)]{Tlatov2009}, and \cite[Chatterjee et al. (2016)]{Chatterjee2016}, from analysis of the Mt Wilson and Kodaikanal series. For the sake of clarity the panels only show the  power spectrum for periods between  2 to 32 years.  Black to  red show low to high values, respectively.}
		\label{fig1}
	\end{center}
\end{figure}

 All studies listed above confirm  some well-known characteristics of past cycles, e.g. they all report that solar cycle 19 showed remarkably high plage coverage and broad latitudinal distribution of active regions. They also agree on the overall increase of the solar activity over the first half of the 20$^{th}$ century and  decrease over the last decades.
However, comparison of results presented in the literature also clearly shows that most of the reported findings  are inherently inconsistent, due to large differences in absolute values and details in the detected short- and long-term trends. These differences affect results derived from distinct time series, as well as findings from the same archives obtained in different studies. 
  
As an example, in Fig. 2 we compare the power spectrum of three published plage area time series   using the Morlet wavelet. In particular, we consider plage area values derived by \cite[Foukal (1996)]{Foukal1996}, \cite[Tlatov et al. (2009)]{Tlatov2009}, and \cite[Chatterjee et al. (2016)]{Chatterjee2016} from the Mt Wilson and Kodaikanal archives and compare them with results for the sunspot number series (SN); focus is on the periods between 2 to 32 years. The dominant feature in all panels is obviously the 11 year cycle, and cycle 19 is the strongest one in all records. However, some differences among the series in the modulation of the high- and low-frequency variations are also seen. It is worth noting that the considered plage area values were independently derived from analysis of different datasets, but the studies by \cite[Tlatov et al. (2009)]{Tlatov2009} and \cite[Chatterjee et al. (2016)]{Chatterjee2016} are based on images from different digitization of the same archive. 

The discrepancy in published results can be mostly attributed to differences in observation set-ups, to different samples of considered data, and to the data processing applied by the various authors.  As shown by  \cite[Chatzistergos et al. (2018b)]{Chatzistergos2018b}, simple visual inspection reveals  considerable variance between the images taken on the same days at different sites, because of instrumental 
and operational characteristics. Furthermore, \cite[Ermolli et al. (2009)]{Ermolli2009} showed that the image quality and contents of three prominent Ca~II~K archives varied with time, probably due to instrument changes, aging of the spectroheliographs used, degradations of the photographic material, and  changes in the observing programs.  
However, another important reason for the discrepancies in published results is the processing applied to the data. Historical observations suffer stronger geometrical distortions  and photometric uncertainties than similar present-day data and hence demand a  
significantly more careful processing than modern observations  to derive meaningful results from their analysis. It is worth noting that most of the results presented in the literature lack error estimates.  \cite[Chatzistergos et al. (2018a)]{Chatzistergos2018a} showed that with proper methods it is possible to derive consistent and accurate results e.g. of the plage coverage for images from different Ca~II~K archives including both historical and modern observations. See also the contribution by 
\cite[Chatzistergos et al. (2018b)]{Chatzistergos2018b} 
in  this volume.

\section{Conclusions}

A large quantity of full-disc Ca~II~K observations have been archived at different sites since the beginning of the 20$^{th}$ century. A few series are still being updated and extended  using either old spectroheliographs or modern telescopes equipped with narrow-band filters. Some historical observations have been digitized and made available to the community. These time series are particularly important because they carry information on the chromospheric magnetic field over a time span exceeding 110 years. Indeed, apart from sunspots, chromospheric plage and network are the next obvious magnetic features on the solar disc and they are clearly spotted  in the Ca~II~K observations. 

Analyses of Ca~II~K time series show both cyclic and long-term variations of plage and network properties. Results presented in the literature from analysis of available data are reasonably in agreement, since they all show some known features of past activity cycles and of long-term solar variation,  but also inherently inconsistent, due to differences in the employed data and the methods used to process the observations.

Understanding variations of both plage and network properties has a broader impact than just describing two of the most prominent solar surface features. However, meaningful results  on the long-term variation of the magnetic field and on solar variability from analysis of Ca~II~K datasets require accurate processing of the available observations and merging of the information stored in  different archives. In fact, analysis of multiple  archives is critical for detecting and correcting instrumental trends, both systematic and otherwise. Besides, study of present-day observations is also  fundamental for optimizing and evaluating the accuracy of the processing applied to historical datasets.


\end{document}